\documentclass[12pt,a4paper]{iopart}

\usepackage{fancybox}
\usepackage{graphicx}
\usepackage{pifont}
\usepackage{amssymb}
\usepackage{pstricks}
\usepackage{times}
\newcommand{\vek}[1]{\mbox{\boldmath $#1$}}
\begin{document}
\title
[Dissociation of H$_2^+$ by short, intense IR pulses] 
{Dissociation spectrum of H$_2^+$  from a short, intense infrared laser pulse: 
 vibration structure and focal volume effects}

\author{ Liang-You Peng, I D Williams and J F McCann}
\address{ International Research Centre for Experimental Physics \\	
School of Mathematics and Physics,\\
                   Queen's University Belfast,\\
                   Belfast BT7 1NN, Northern Ireland, UK. \\
                   \vskip0.2cm
                   email : l.peng@qub.ac.uk}
\begin{abstract}
The dissociation spectrum of the hydrogen molecular ion  by short intense 
pulses of infrared light is calculated.  The time-dependent 
Schr\"odinger equation is discretized and integrated 
in  position and momentum space. For few-cycle 
pulses one can resolve  vibrational structure that 
commonly arises in the experimental 
preparation of the molecular ion from the neutral molecule. 
We calculate the corresponding energy spectrum and analyze the dependence  
on the pulse time-delay,  pulse length, and intensity of the  laser 
for $\lambda \sim 790$nm.
We conclude that the proton spectrum is a both a sensitive probe of the
vibrational dynamics and the laser pulse. Finally we compare our results 
with recent  measurements of the proton spectrum  for 55 fs pulses using 
a Ti:Sapphire laser ($\lambda \sim 790 $nm). 
Integrating over the laser focal volume, for the   
intensity $I \sim 3 \times 10^{15}$W cm$^{-2}$, we find 
our results are in excellent agreement with these experiments.

\end{abstract}
{\it To be submitted to J. Phys. B} 

\section{Introduction}

Energy transfer mechanisms in  molecules exposed to short 
intense laser pulses is of great current interest \cite{Post01}.
Processes such as multiple photoionization, multiphoton dissociation
and high-order harmonic generation have been studied 
for a wide-range of molecular species and  an excellent up-to-date review 
of the field has been provided by Posthumus  \cite{Post04}.
 The hydrogen molecule and 
molecular ion represent ideal systems for 
a detailed understanding of ionization and dissociation dynamics of small
molecules \cite{Fras93}. 
The simplicity of molecule means that it is the system 
of choice for theoretical studies. In general the mechanisms of 
 dissociation \cite{Gius95} and ionization \cite{Plum95,Peng03,Peng04}
are very well characterised for this molecule. Nonetheless, the 
solution of  three-body disintegration, 
incorporating spatial and temporal variations in the laser pulse
as well as the thermal ensemble of molecular states, remains a severe challenge 
for simulation. Consequently, progress has been fairly limited in 
dissociative ionization spectrum calculations.

Although there exists a wealth of data on the photofragment energy 
spectrum for the neutral molecule, H$_2$, and its deuterated forms
\cite{Post04}.  the data for the much simpler isolated molecular 
ion is extremely scarce. The feasibility of experimental 
studies is hampered by the difficulty in preparing
the molecular ion. In fact experiments on   H$_2^+$  in intense laser fields has 
only become possible in the last 4 years \cite{Will00, New03, Sand00,Pavi03} due 
to refinements in ion sources and charged molecular beam spectroscopy. 
Typically a small portion of the molecules, at a temperature 
of a few hundred Kelvin and hence predominantly in the     
vibrational state $(v=0)$, can be converted into the bound molecular ion  by 
electron-impact ionization,
\begin{equation}
e^- + {\rm H}_2(^1\Sigma_g, v=0)  \rightarrow e^-+e^- + {\rm H}^+_2(^2\Sigma_g, v') \ \ .
\label{coll}
\end{equation}
The molecular ions can be extracted, cooled and collimated into a beam that 
can be injected to the focus of the laser. The photodissociation and/or 
photoionization processes, 
\begin{equation}
n'h\nu + {\rm H}^+_2(^2\Sigma_g, v') \rightarrow {\rm H}^*+{\rm H}^+ +n''h\nu
\label{photo}
\end{equation}
produce ions  and electrons that can be collected and analysed. 
Owing to the small nuclear mass, the expansion (vibration) is very rapid - on
the scale of 10 fs or less. Consequently the molecule relaxes 
extremely rapidly on the timescale of the pulse 
rise time. The design of infrared pulses that would, among other things, allow 
the resolution of  timescale of a fraction of a  femtosecond.

The main aim of this paper is a direct comparison of theory and experiment for the 
proton energy spectrum produced by a short intense infrared pulse interacting 
with H$_2^+$ \cite{Will00,Sand00}. 
Fragment-ion coincidence measurements that isolate
the dissociation and dissociative ionization processes and make the  
comparison of theory and experiment feasible 
and realistic. The molecular dissociation  
dynamics are modelled using a  two-state approximation and 
takes account of the laser pulse profile. 
Although this approach neglects the ionization channels, 
the results for dissociation energies 
are in excellent agreement with the observations.

\section{The model}
\subsection{Physical considerations}

The characteristic vibrational and rotational time of the 
H$_2^+$ molecule are $T_{\rm vib}\sim 15$ fs and 
$T_{\rm rot} \sim 170$ fs, respectively.
For a pulse duration 60 fs or less, the laser interaction 
is sudden on the rotational timescale. Thus one 
can assume a random statistical 
orientation of the molecular axis, and that the nuclei 
recoil along the axial direction. We assume  that the  
bound ion is created by vertical transitions from the 
$v=0$ state of the neutral $H_2 (X^1\Sigma_g)$.  
The parameters of this potential surface are,   
bond length $R_0 = 1.40$a.u. with
 vibrational constants, $\hbar\omega_0= 0.546$eV and $x_e=0.0276$. 
 The $v=0$ wavepacket is promoted onto 
the corresponding molecular-ion 
surface  $H_2^+ (X^2\Sigma_g)$ with 
parameters $R_0'=2$a.u., $D_0'=2.65$eV, 
$\hbar\omega'= 0.288$eV and $x'_e=0.0285$  \cite{Hube79}. 
The $v=0$ state is projected onto the 
$v'$ manifold of states according to the Franck-Condon 
principle \cite{Dunn66}.  The manifold of $v'$-states 
is then subjected to intense Ti:Sapphire light. 


In the dissociation process, momentum is conserved, for a symmetric pulse, and the recoiling 
atom and ion have equal and opposite velocity in the centre-of-mass frame. 
The ions and atoms are collected in coincidence such that dissociation 
and dissociative ionization  can be discriminated. However, 
depending on the method of measurement, the atomic fragments 
can arise
from the entire focal volume of the laser.  So a valid theoretical 
comparison must firstly establish the single molecule 
energy spectra, and then sum these spectra with the 
appropriate focal volume weighting for the experiment. 
The intensity  $I_0(\rho,z)$  of a focused cylindrically-symmetric 
 laser beam is spatially 
Gaussian in the radial $r$-direction 
and Lorentzian along the axis $z$ \cite{Post01}
\begin{equation}
I_0(\rho,z) = \frac{I_f}{ 1+ \left( z/z_R\right)^2} \exp\left\{\frac{-2
\rho^2}
{w_0^2 \left[1+ \left( z/z_R\right)^2\right]}\right\}.
\end{equation} 
where the minimum waist of the beam $w_0$ and Rayleigh range $z_R$ are given by 
\begin{equation}
w_0 = \frac{2 f \lambda }{\pi D}, \qquad {\rm and} \qquad
z_R = \frac{\pi w_0^2 }{\lambda}
\end{equation}
respectively.  In our method, 
we first discuss the single-molecule response in detail, 
and then describe the orientation-averaged 
and focal-volume integrated results. 
 
The electronic coordinate, with respect to the origin at the internuclear 
midpoint, is denoted by $\vek{r}$ and the internuclear coordinate 
is written as $R$. The two lowest electronic states, namely the 
ground state $X^2\Sigma_g^+ \ \  
(\phi_g)$, and   $A^2\Sigma_u^+ \ \ (\phi_u)$ 
are sufficient for the study of pure dissociation dynamics. 
The corresponding adiabatic energies are $E_g(R)$ and 
$E_u(R)$.  Then the two-state approximation is
\begin{equation} 
\Psi(\vek{r}, R, t) = F_g(R,t)\phi_g(\vek{r}, R)  + F_u(R,t)
\phi_u(\vek{r}, R)
\label{20040224a}
\end{equation}
where $F_g(R,t)$ and $F_u(R,t)$ are the time-dependent nuclear 
wavefunctions. In restricting the electronic spectrum to 
the two lowest levels, the calculations will inevitably be 
gauge dependence. For low frequency fields (much less than 
the energy level gap) it is essential to use the length gauge for 
the laser-molecule interaction. Let us take the light as linearly 
polarized along the direction $\vek{\varepsilon}$ , and  denote the 
dipole moment between $\phi_{u,g}$ as $\mu(R)$. Writing  the 
electric field as $E(t)$, the coupling potential is denoted by 
 $V_L(R,t)= -\mu(R)E(t) u$, where 
 $ u \equiv \vek{\varepsilon} \cdot \vek{\hat{R}}$.
Within the axial-recoil approximation,  
$\theta_k=\cos^{-1}u$, is the angle of ejection 
of the ion(atom) with respect 
to the polarization vector. 
Averaging over the 
molecular orientation is equivalent to  averaging  the projection 
of the electric field along the molecular axis. Since
the dissociation rate increases rapidly with intensity 
over the range, $I \sim 10^{12}-10^{14}$ W cm$^{-2}$, it follows 
that the atoms and ions  are ejected predominantly along the polarization 
direction. It is highly anisotropic. However at the higher 
intensities, $I \sim 10^{15}$ W cm$^{-2}$, the dissociation process 
begins to saturate and, as we will show, the angular distribution 
is broader.


\subsection{Numerical method}

Taking $m={\textstyle {1 \over 2}} m_p$ to denote 
the reduced mass of the protons, and using atomic 
units, the two-state coupled  equations are:
\numparts
\begin{eqnarray}
\fl \ \ \ \ \ \ i\frac{\partial}{\partial t}F_g(R,t)&=&-\frac{1}{2m}
\frac{\partial^2}{\partial R^2}F_g(R,t)+E_g(R)F_g(R,t)
+V_L(R,t)F_u(R,t), 
\label{c1} \\
\fl \ \ \ \ \ \ i\frac{\partial }{\partial t}F_u(R,t)&=&-\frac{1}{2m}
\frac{\partial^2}{\partial R^2}F_u(R,t)+E_u(R)F_u(R,t)
+V_L(R,t) F_g(R,t), 
\label{c2}
\end{eqnarray}
\endnumparts
To a very good approximation \cite{Bunk73} the potential functions can 
be written in the form,
\begin{eqnarray}
E_g(R) &=& 0.1025 \left[e^{-1.44(R-2)} -2e^{-0.72(R-2)} \right],  \\
E_u(R) &=& 0.1025 \left[e^{-1.44(R-2)} +2.22e^{-0.72(R-2)} \right] 
\label{morse}
\end{eqnarray}
Similarly, the dipole moment \cite{Bate51} can be fitted by the function \cite{Lin01}
\begin{equation} 
\mu(R) = -\frac{1}{(2+1.4R)} + \frac{R}{2\sqrt{1-p^2}}
\end{equation}
with,  $p=(1+R+R^2/3)e^{-R}$. For a frequency $\omega_L$ and peak field 
strength $E_0$. This is related to the cycle-average 
intensity, $ I ={\textstyle {1 \over 2}}c \varepsilon_0 E_0^2$.
The time-dependence of the field can be written explicitly as 
\begin{equation}
 E(t) = E_0 f(t) \cos \omega_L t
\end{equation}
In this paper we use the Gaussian profile that most closely models 
the pulses in the experiment of interest,
\begin{equation}
 f(t) = \exp \left[- (4\ln2)\  \left( \frac{t-T_c}{T_p} \right)^2 \right]
\end{equation}
where $f(T_c)=1$ is the maximum, and $T_p$ defines the  duration of the 
laser pulse. 
\subsection{The grids}
Discretization methods previously 
developed for photodissociation \cite{Joli92} and 
photoionization of molecules \cite{Dund03,  Peng03, Peng04}
can be readily applied.  
The discrete-variable representation (DVR) has proven to be a very efficient and 
accurate method in solving both time-independent and time-dependent 
Schr\"{o}dinger equations. 
The Cartesian Lagrange functions \cite{Baye86} are given by 
\begin{eqnarray}
f_i(x)& =&
\sum\limits_{k}\varphi_k^\ast(x_i)\varphi_k(x)
 \nonumber \\
      & =& \frac{1}{N}\sum\limits_{k}\exp\left[-i\frac{2\pi
       k}{N}x_i\right]\exp\left[i\frac{2\pi k}{N}x\right]  
\label{c/08/02/04}
\end{eqnarray}
with the mesh points $x_i = i- {(N+1)}/{2}$ ($i=1,2,...,N$).  
The radial coordinate $R$ is 
discretised on this mesh  Cartesian mesh
 between the limits $R_{\rm min} \leq R_j \leq 
R_{\rm max}$, with $j \in \{1,2,3, \dots, N-1,N\}$, such that:
\begin{equation}
R_j = {(R_{max}-R_{min})\over N-1}
\left[ j- {\textstyle {1 \over 2}}(N+1) \right] +
 {\textstyle {1 \over 2}} (R_{max}+R_{min}) 
\end{equation}
Then the wavefunctions are expanded as:
\begin{eqnarray}
F_{\sigma}(R,t)&=&\sum\limits_{i=1}^{N} F_{\sigma}(R_i,t) f_i(R)\nonumber \\
&=& \frac{1}{N}\sum\limits_{i=1}^{N} F_{\sigma}(R_i,t)
\sum\limits_{m=1}^{N} \exp{\left[i\frac{2\pi x_m}{N}\frac{(R-R_i)}{h}\right]} 
\hspace*{0.2cm}(\sigma = g, u)
\label{psiexpan}
\end{eqnarray}
where $h=(R_{max}-R_{min})/(N-1)$.
Then the matrix equation corresponding to equations (\ref{c1} and \ref{c2})
is a dense set of $2N$ 
linear equations ($\sigma=g,u$):
\begin{equation}
{\displaystyle 
\sum_{j=1}^N  T_{ij} F_{\sigma, j}(t) + \sum_{\tau,j}V_{\sigma \tau, ij}(t)
 F_{\tau, i}(t)
=i\dot{F}_{\sigma, j}(t)
}
\end{equation}
The  matrix elements of the kinetic operator  are given 
by \cite{Baye86}
\begin{equation}
T_{ij} = \left\{ 
 \begin{array}{lr}
   \frac{\alpha\pi^2}{6} \left( 1-\frac{1}{N^2}\right)& i=j  \\
 (-1)^{i-j}\frac{\alpha\pi^2}{ N^2}\frac{\cos\left[\pi(i-j)/N\right]}
   {\sin^2\left[\pi(i-j)/N\right]} & i\neq j.
  \end{array}
 \right.
 \end{equation}
where the scale factor is, $\alpha=(N-1)^2 m^{-1}(R_{max}-R_{min})^{-2}$. 
The diagonal potentials are:
\begin{equation}
V_{uu, ij}(t)=  \delta_{ij} E_{u}(R_j)
\ \ \ \ \ \ \ \ \
V_{gg, ij}(t)=  \delta_{ij} E_{g}(R_j)
\end{equation}
with the off-diagonal coupling:
$
V_{ug, ij}(t)= V_{gu, ij}(t)=
 \delta_{ij}  V_L(R_j,t)
$
 We integrate 
the  differential equations  using the 18th order
Arnoldi propagator as described by Peng et al \cite{Peng04}.

As is well known \cite{Bate51}, for $R \rightarrow \infty$, the moment 
$\mu(R) \rightarrow {\textstyle {1 \over 2}R}$, and the coupling is divergent in 
the molecular basis. This is simply a manifestation of the molecular 
basis and the use of the length gauge.
It is necessary and convenient to transform to the
asymptotic decoupled atomic eigenstates for the energy spectrum. 
These asymptotic states then evolve adiabatically at 
large distances and long times. However  these 
states in turn are unbounded in configuration space. 
As shown by Keller \cite{Kell95}, it is possible to project 
the diffuse adiabatic asymptotic states onto a compact momentum 
space.  More importantly, this methods produces the energy spectrum of the atoms 
that can be compared with experiment.

Following  \cite{Heat87,Kell95}  
we divide the entire range of $R$ into two regions; an internal region({\it in}), 
where the molecular forces are significant, and an 
asymptotic region ({\it as}) where they are not. The short-range polarization
potential means that the interaction region can be comparatively small 
(of the order 20 a.u.). Then the  wavefunction can be partitioned as follows:
\begin{equation}
F_{\sigma}(R,t)= F^{in}_{\sigma}(R,t)+F_{\sigma}^{as}(R,t)
\end{equation}
where:
\begin{eqnarray}
F_{\sigma}^{as} (R,t) &=& (1-S(R)) F_{\sigma} (R,t) \\ \nonumber 
F_{\sigma}^{in}(R,t) &=& S(R) F_{\sigma}(R,t) \hspace*{1cm} ({\sigma}=1, 2). 
\label{splitwave}
\end{eqnarray}
with  
\begin{equation}
S(R) = \left( 1 + \exp\left[ \frac{R-R_s}{\tau_s}\right]\right)^{-1} 
\end{equation}
where $R_s$ is dividing point and $\tau_s$ is a parameter which determines the 
smoothness of the partition. 
In the asymptotic region Hamiltonian  is diagonalised by 
transforming to the atom+ion states $\chi_{1,2}$:
\begin{equation}
\left( 
\begin{array}{c}
\chi^{as}_1(R, t) \\
\chi^{as}_2(R, t)
\end{array}
\right)  
= {1 \over \sqrt{2}} 
\left( 
\begin{array}{cr}
 1 &  1 \\
 -1 & 1 
\end{array}
\right) 
\left( 
\begin{array}{c}
F_g^{as}(R,t) \\
F_u^{as}(R,t)
\end{array}
\right).  
\end{equation} 
These states evolve into 
superpositions (sums over momentum) of the  asymptotic 
states:
\begin{equation}
\fl {\chi}^{\pm}_k(R,t,t')=
{1\over \sqrt{2\pi}}
\exp\left[ i(k\mp \Delta(t,t')) R  
-i\frac{1}{2\mu}\int_{t'}^t d\tau
[k\mp\Delta_0(t,\tau)]^2\right] 
\label{scatt}
\end{equation}
where the ion quiver momentum is
\begin{equation}
\Delta_0(t,t')= \frac{1}{2}\int_{t'}^td\tau' E(\tau')
\end{equation}
A symmetric or zero-area pulse is such that: $\Delta_0(+\infty,-\infty)=0$.

\subsection{Transformation to momentum space}

The key step in obtaining an energy spectrum is the 
projection of the numerical wavefunction onto the  the continuum of asymptotic  states 
(\ref{scatt}). It is efficient to do this by discretizing the $k$-space and performing 
a finite Fourier transform \cite{Kell95}. 
One difficulty with such an approach is that the momentum shift 
$\Delta_0$ is a continuous 
function and inevitably leads to a mismatch of the $k$-grid. 
In the present calculations,  we use the quadrature rule for the 
DVR to calculate the Fourier transform directly.  We define the 
shifted Fourier transform:
\begin{eqnarray}
 \hat{\chi}_{1,2}(k&\mp&\Delta_0(t,t^{\prime}),t)= 
 \frac{1}{\sqrt{2\pi}} \int_{R_{min}}^{R_{max}} e^{-i(k\mp\Delta_0(t,t^{\prime})R} 
 {\chi}_{1, 2}(R,t) dR \nonumber \\
 &=&  \frac{1}{ N\sqrt{2\pi} } \sum\limits_{i=1}^{N} \chi_{1,2}(R_i,t)\sum\limits_{m=1}^{N} 
 \frac{e^{-i 2 \pi x_m R_i / Nh}}{-i\left[ (k\mp\Delta_0(t,t^{\prime})) - 
 2 \pi x_m /Nh\right]} \nonumber \\
   & \times&\left\{ \exp\left[ -i R_{max} \left( (k\mp\Delta_0(t,t^{\prime})) -
 \frac{2\pi
 x_m}{Nh}\right)\right] \right.  \nonumber \\
 &-& \left. \exp\left[ -i R_{min} \left(
 (k\mp\Delta_0(t,t^{\prime})) -\frac{2\pi
 x_m}{Nh}\right)\right]
 \right\}.
 \label{myprop}
 \end{eqnarray}
Although this quadrature is less efficient than the fast-Fourier transform, 
it allows for much greater flexibility in choosing the $k$ and $R$ grids.

At each time step, the projection onto the momentum space is made and 
the  coherent momentum space wavefunction is calculated. This process is 
repeated and continued for some time after the end of the pulse. This 
allows the dissociating wavepacket still within the interaction region 
time to reach the asymptotic zone, and for the low energy 
components to be captured. Typically we extend the integration 
time to $4 T_c$ for this purpose.
The $k$-space probability density is then given by \cite{Kell95}:
\begin{equation}
P_k(k) = \lim_{t \rightarrow \infty} \vert \hat{\chi}_1^{as}(k,t)+\hat{\chi}_2^{as}(k,t) \vert^2
\end{equation} 
This choice of normalization gives:
\begin{equation}
P_d = \int_0^{+\infty} P_k(k)\ dk  
\end{equation} 
is the total probability for dissociation. The dissociation 
energy is shared equally by  the fragments so that the proton energy 
is given by $E_p= (4m)^{-1}k^2$.  
The energy spectral density can then  be calculated from the equation: 
\begin{equation}
P_E(E_p)= (dk/dE_p)P_k(k) = (m_p/k) P_k(k) . 
\end{equation}

\section{Results and discussion}

\subsection{Dissociation from  H$_2^+\ (v'=0)$ at $\lambda=790$nm}

\begin{figure}[t]
\centering
\includegraphics[clip=true,width=13.5cm]{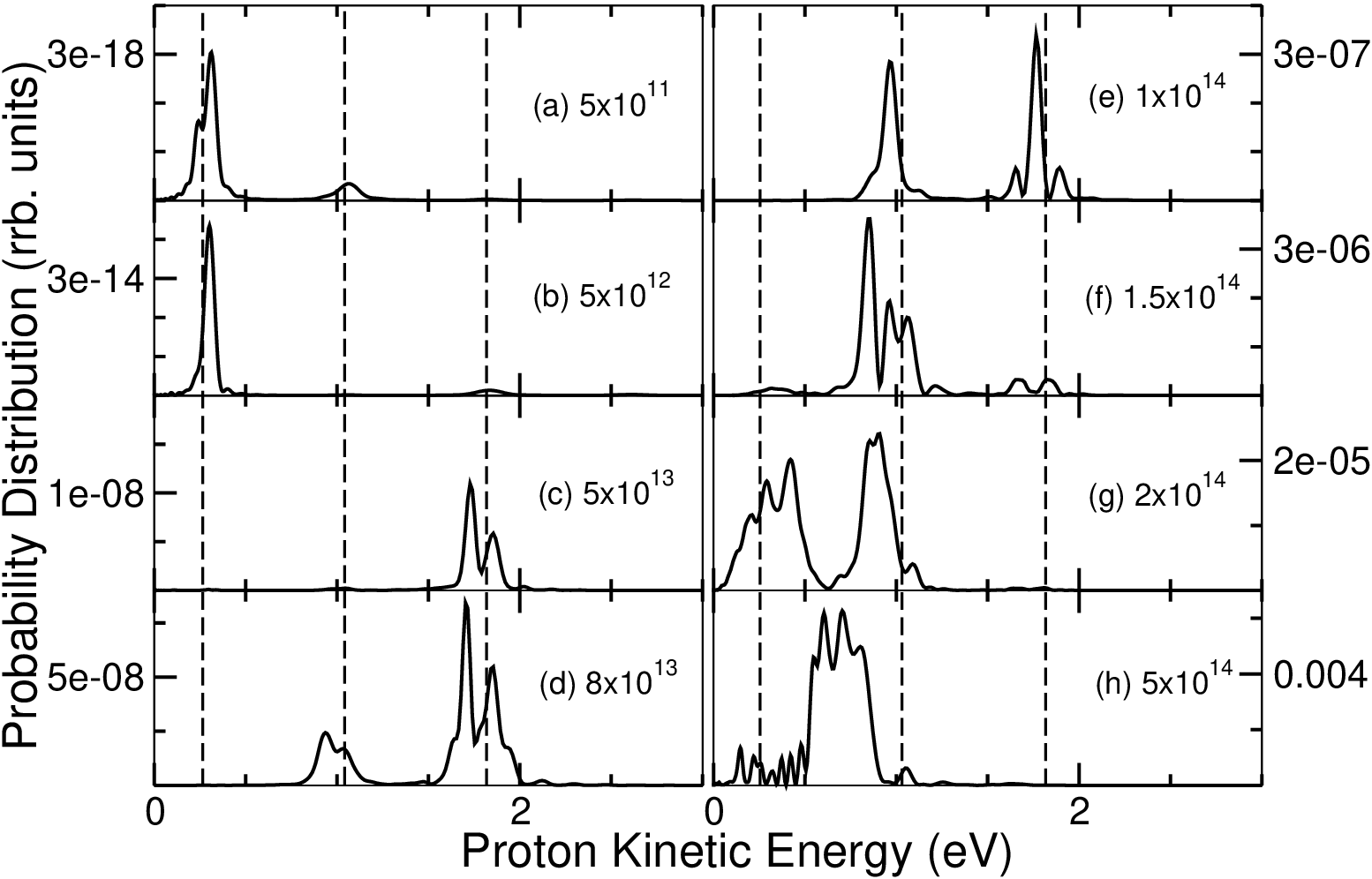}
\caption{
Proton  energy spectrum $P_E(E_p)$ for  H$_2^+ (v'=0)$ with 
$\lambda=790$nm and pulse duration $T_p=55$fs. The vertical lines 
indicate the anticipated energy of the two-photon, three-photon 
and four-photon absorption from $v'=0$.} 
\label{figure1}
\end{figure}

Firstly we consider photodissociation directly
from $v'=0$ of H$_2^+$. There are several very accurate 
proton spectra  calculation for 
$v'=0$ dissociation by $\lambda \sim 330$nm pulses 
for  $I \sim  10^{12}-10^{14}$ W cm$^{-2}$ 
\cite{Miret94,Kell95}. We checked our calculations with 
these results and found excellent agreement in all cases. 
Now consider long-wavelength dissociation, we 
recall that the dissociation energy of the $v'=0$ state is
$D_0'=2.65$ eV.  For $\lambda \sim 790$nm, corresponding to the Ti:Sapphire 
laser, the photon energy is 1.57 eV. Therefore dissociation 
is necessarily a second or higher order  process.
In figure \ref{figure1} we present results for the proton 
energy spectrum $P_E(E_p)$ for a $T_p=55$fs pulse for a range of 
intensities.
In these calculations, the number of points in $R$-space $N=512$ and the number of points 
in momentum space $N_k = 2048$, with   $R_{min} = 0.1$ a.u. and $R_{max} = 28.5$ a.u..
For the splitting procedure we use, 
$R_s \sim 0.7 R_{max}$ and $\tau_s=0.2$. 
For the time-dependent propagation we use 
$\delta t = 0.01 $.

For a pulse of this length, the bandwidth (FWHM) is narrow  $\sim 0.03$ eV, 
and this is reflected in the sharply-defined proton energy  (figures 1a,1b and 1c)
in the perturbation regime.  
In figure \ref{figure1}, the  relative strength of the  two-, three- and 
four-photon dissociation channels change dramatically with the variation of the 
pulse intensity. There is a propensity for the high-order process. 
As the interaction increases there is 
a   leftward shift of the three-photon peaks
in frame (e) to (g). This is due to the downwards Stark shift of the ground state. 
The leading-order  term of this shift is approximately  linear 
with  intensity. 
The broadening of the peaks is evidence of the 
mixing of vibrational levels and shortening of the lifetime of the decaying 
state $v'=0$. In the last figure (h) at 
the highest intensity the spectrum is dominated by a large broad peak 
around 0.7 eV where the field-free vibrational structure is destroyed. 
At intensities above $ 10^{15}$ W cm$^{-2}$, the process 
is saturated  by tunelling dissociation  and the spectrum is dominated by low energy 
(less than 0.5 eV) protons \cite{Gius95}. This data underlines  
the difficulty in using time-independent perturbation theory to determine 
the ion spectrum and yield under these conditions.
 
\begin{figure}[t] 
\centering
\includegraphics[clip=true,width=13.5cm]{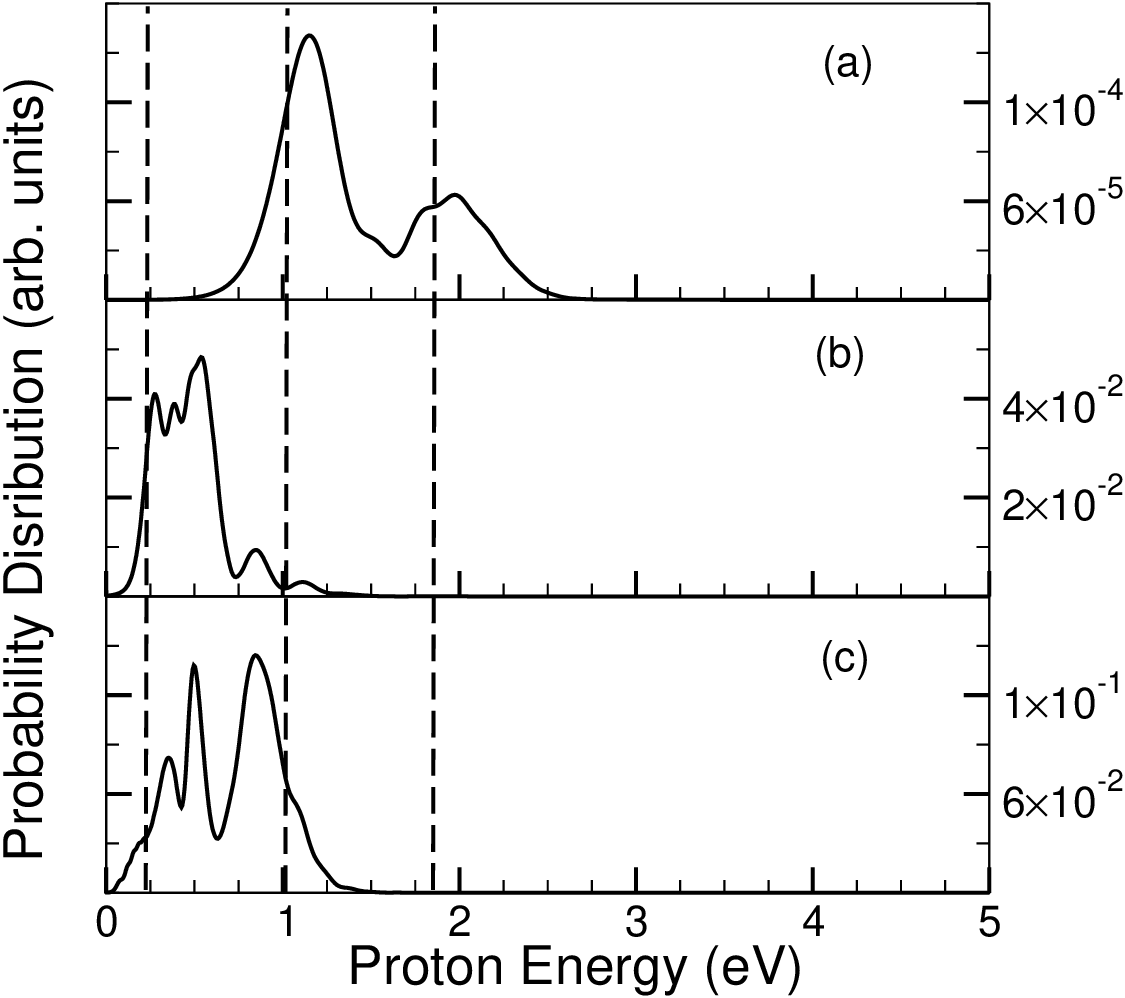}
\caption{
Proton  energy spectrum $P_E(E_p)$ for  H$_2^+ (v'=0)$ with 
$\lambda=790$nm and I=$5.0 \times 10^{14}$ W cm$^{-2}$. The figure 
shows the effect of increasing pulse length 
 (a).$T_p = 5$ fs; (b). $T_p = 10$ fs; (c). $T_p = 20$ fs.
 The three vertical lines indicate the  weak-field 
 two-photon, three-photon and four-photon absorption energies.} 
\label{figure2}
\end{figure}

The second calculation (figure \ref{figure2}) concentrates on the 
effect of pulse length for a fixed  intensity 
$I=5\times 10^{14}$ W cm$^{-2}$. For  pulses shorter 
than 15 fs  the vibrational phase is accepted to play a role. 
However it is also important to note that the bandwidth has a significant 
effect. For example, a 5 fs pulse has a  bandwidth of the 
order  0.35 eV. This would explain the skewness of the peak in 
figure 2(a), and the presence of the 4-photon peak. 
However, in figures 2(b) and 2(c), the results resemble the narrow-band
data in figure 1. The oscillations in  the peaks reflect 
wavepacket oscillations within the well and we 
discuss  this in more detail later when investigating 
dissociation from excited vibrational 
states.

\subsection{Dissociation from H$_2\ (v=0)$ at $\lambda=790$nm}

\begin{figure}
\centering
\includegraphics[clip=true,width=10.0cm]{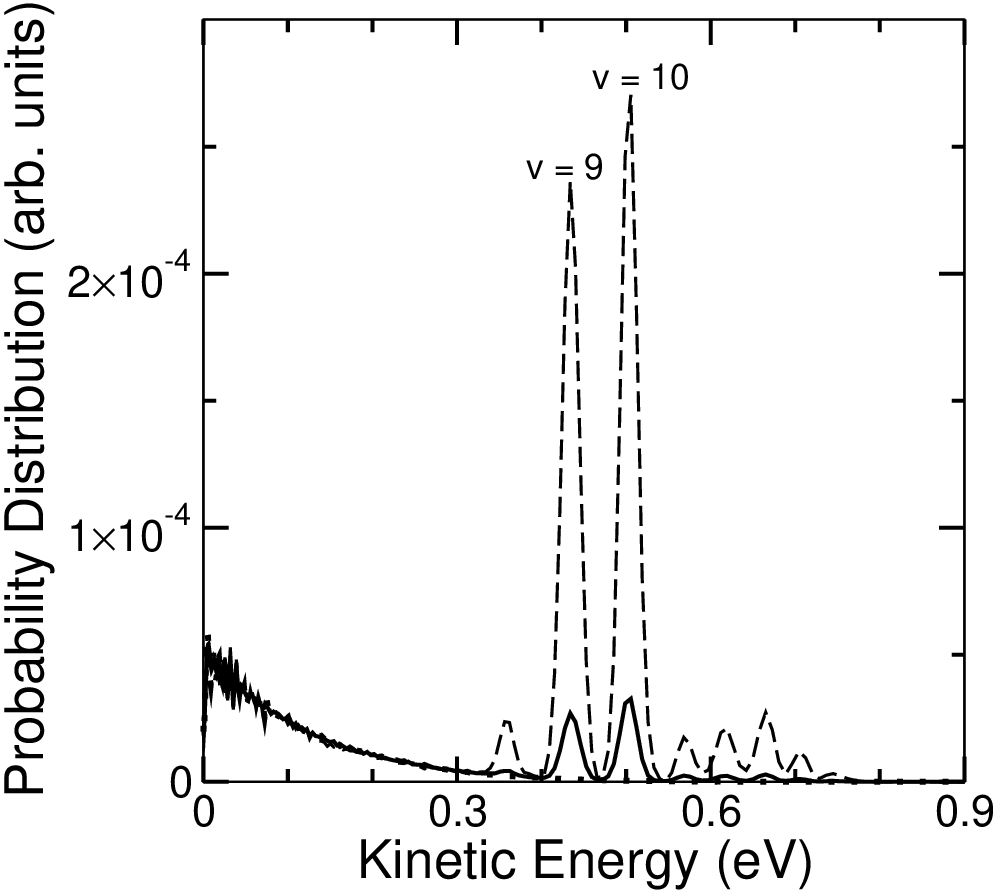}
\caption{Proton energy distribution $P_k(E_k)$ at low laser intensities.
The pulse length is $T_p = 55$ fs and  $\lambda=790$ nm. 
Full line: I=$5\times 10^{9}$ W cm$^{-2}$; 
dashed line: I=$5\times 10^{10}$ W cm$^{-2}$. 
The low-energy background   (dotted line) is 
the shake-off spectrum. The  
vibrational lines are single-photon peaks. The 
high intensity lines are 10 times stronger 
than the low intensity results as expected from perturbation theory.} 
\label{figure3}
\end{figure}

In the ion-beam experiments,   H$_2^+$ 
 is prepared from the collisional ionization process 
(equation \ref{coll}). The H$_2\ (v=0)$ 
state is projected onto a coherent superposition 
of the H$_2^+$ vibrational states, $F_{v'}$, 
where $v'=0,1, \dots,17$. The   
amplitudes $C_{v'}$ are given by  the Franck-Condon factors. So 
that 
\cite{Dunn66} 
in the absence of the laser the wavepackt evolves coherently
\begin{equation}
F_g(R,t') = \e^{i\phi(t)}\sum_{v'=0}^{17} C_{v'}e^{i\phi_{v'}-iE_{v'}t'} F_{v'}(R)
\end{equation}
where the relative phases $\phi_{v'}$ are well defined constants, 
and $E_{v'}$ are the vibrational energies. 
The overall phase $\phi(t)$ depends on the time of formation of the ion.
According to the Franck-Condon factors,
the largest coefficients correspond to $v'=2,3,4$. However  
a small fraction  of the population 
will be above the dissociation limit
leading to shake-off dissociation. 
Dunn's \cite{Dunn66} estimate of this population was 
around 2.8\%. In the present calculations, using the energy 
curves of equation (\ref{morse}), we obtain a value  around 0.8\%.
In experiment the shake-off process is eliminated and plays no role. However, 
the coherence of the $v'$ states has an important role.
Since $t'=0$ is determined by the start of the pulse 
the molecules arrive in the focal volume with random vibrational phase.  
The random nature of the time delay is equivalent 
to averaging over random vibration phase.
We can readily illustrate the physical consequences of this time delay 
in the proton spectrum.
Suppose we choose the initial wavepacket as: 
$$
F_g(R,t'=0) = F_{H2}(v=0,t=0)
$$
Then 0.8\% of this wave function will naturally dissociative (shake-off) without   
laser interaction. 
In the calculations 
shown in \ref{figure3}, we present the proton energy 
distribution  in  the perturbation regime for 
a long pulse. The laser intensities are $5\times10^{9}$ W cm$^{-2}$ and
 $5\times 10^{10}$ W cm$^{-2}$
respectively, with  $\lambda = 790 $ nm and pulse duration 
$T_p = 55$ fs. The intensity-independent background that 
peaks at zero energy is the shake-off process. The rapid oscillations 
in this background are artifacts arising from the numerical method. 
The lowest energy states emerge most slowly and require long 
propagation times. The major source of noise is the splitting procedure 
that creates an interference pattern in the long-wavelength wavepacket 
components. However, above 0.3 eV the interference artifacts disappear
and we are confident in the 
numerical accuracy of the energy spectrum in this range.

The 1-photon dissociation threshold means that only levels $v' \geq 5$ 
appear. However the peak heights are modulated 
according to the values of both  $C_{v'}$ and the Franck-Condon dissociation 
factors. This vibrational structure has been resolved and 
measured experimentally \cite{Sand00,Pavi03}.  
As predicted by the perturbation theory, the probability for the individual peak 
is proportional to the laser intensity.
Very recently, Serov and coworkers \cite{Sero03} have obtained results 
of excellent agreement with these experimental measurements after  careful 
consideration of averaging over initial ro-vibrational states and over the 
focus volume effect of the laser pulse.

\begin{figure}[t]
\centering
\includegraphics[clip=true,width=12.5cm]{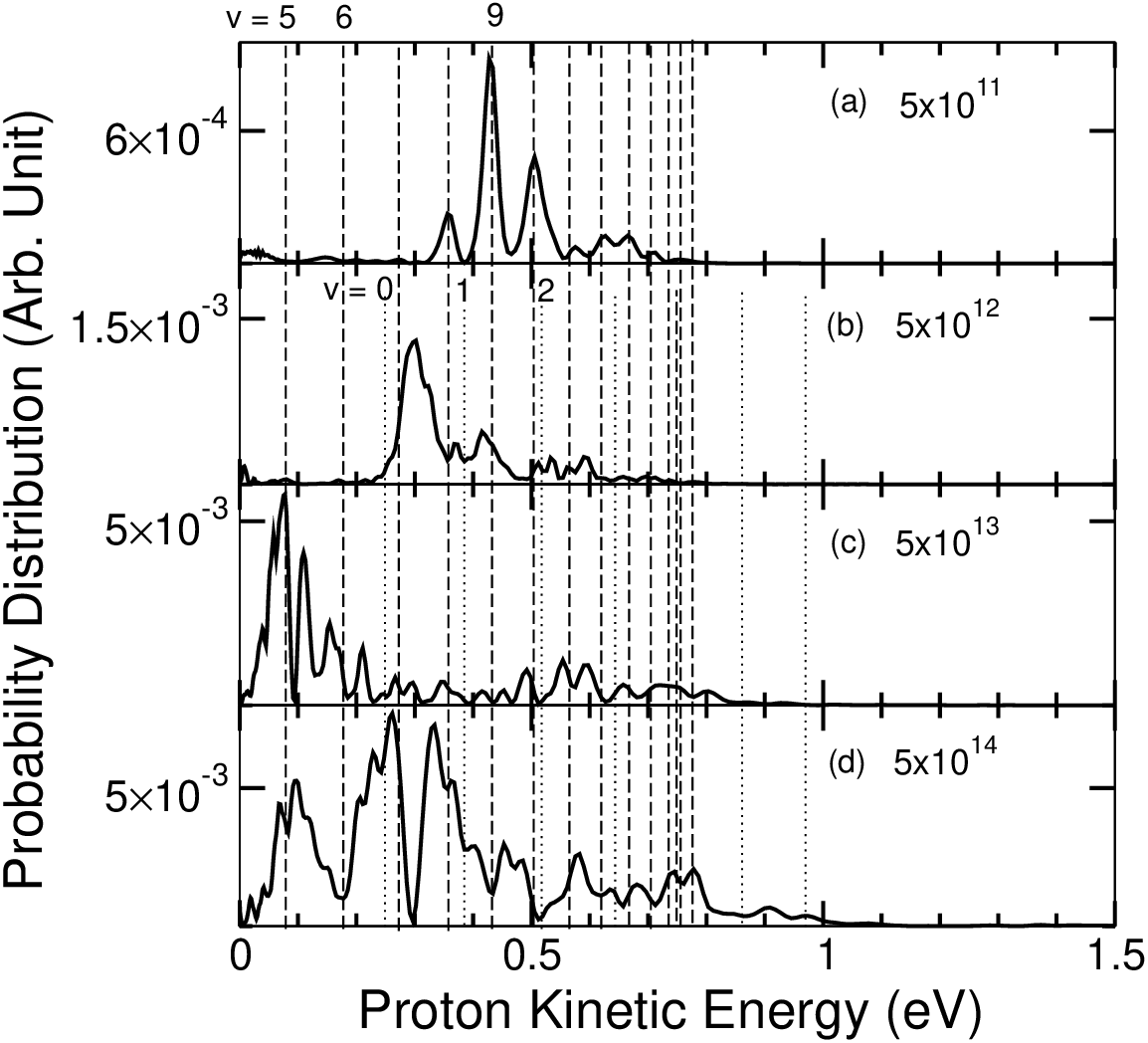}
\caption{Proton  energy distribution  at different intensities. 
The initial vibrational wavepacket is prepared  from H$_2$ $(v=0)$. 
. The pulse parameter are $T_p = 55$ fs and $\lambda=790$ nm. The dashed
vertical lines indicate the one-photon vibrational  
release with one photon absorption from different vibrational states and the 
dotted vertical line show those with two photon absorption.} 
\label{figure4}
\end{figure}

In figure (\ref{figure4}) the results for high intensity  show 
dramatic differences with perturbation theory. 
In order to compare with experiment, it is desirable  to eliminate the 
spurious shake-off background. This can be achieved by projecting out the continuum
or by waiting for the shake-off to occur and absorb the dissociation. 
In fact, the shake-off background is an imperceptible  perturbation to the results
at high intensity, as shown in figure \ref{figure4}. In practice, 
for intensities above $5 \times 10^{11}$ W cm$^{-2}$, we have 
not found it necessary to remove the shake-off wavepacket a priori. 
At higher intensities the two-photon dissociation process 
becomes important, and in figure \ref{figure4} (b) we include vertical dotted 
lines to indicate the expected positions of these peaks. 
Again, at the highest intensities, the spectrum peaks move towards 
the lower energies although the spectrum also broadens. 
The peak heights in figure \ref{figure4} (c) and (d) are 
similar, but the yield is much higher for the higher intensity.
The corresponding dissociation probabilities $P_d$ for 
each intensity  are ????.
As mentioned before, increasing the angle between molecule and polarization 
is equivalent to decreasing the intensity. Then figure \ref{figure4} 
represents the angular-differential dissociation rates. Thus for a spatially 
uniform pulse of cycle-average intensity $I=5\times 10^{14}$ W cm$^{-2}$, 
figures \ref{figure4} (b),(c) and (d) are respectively the 
yields at $\theta=$72, 84    and  88  degrees.  For pulses 
much longer than the vibrational time,  the initial phase of the molecular wavepacket 
is not a significant factor. Averaging over the initial 
phase has little effect. In the next section we will consider 
pulses of duration comparable to the vibrational dephasing time.

\subsection{Vibrational phase and pulse length}

\begin{figure}
\centering
\includegraphics[clip=true,width=10.0cm]{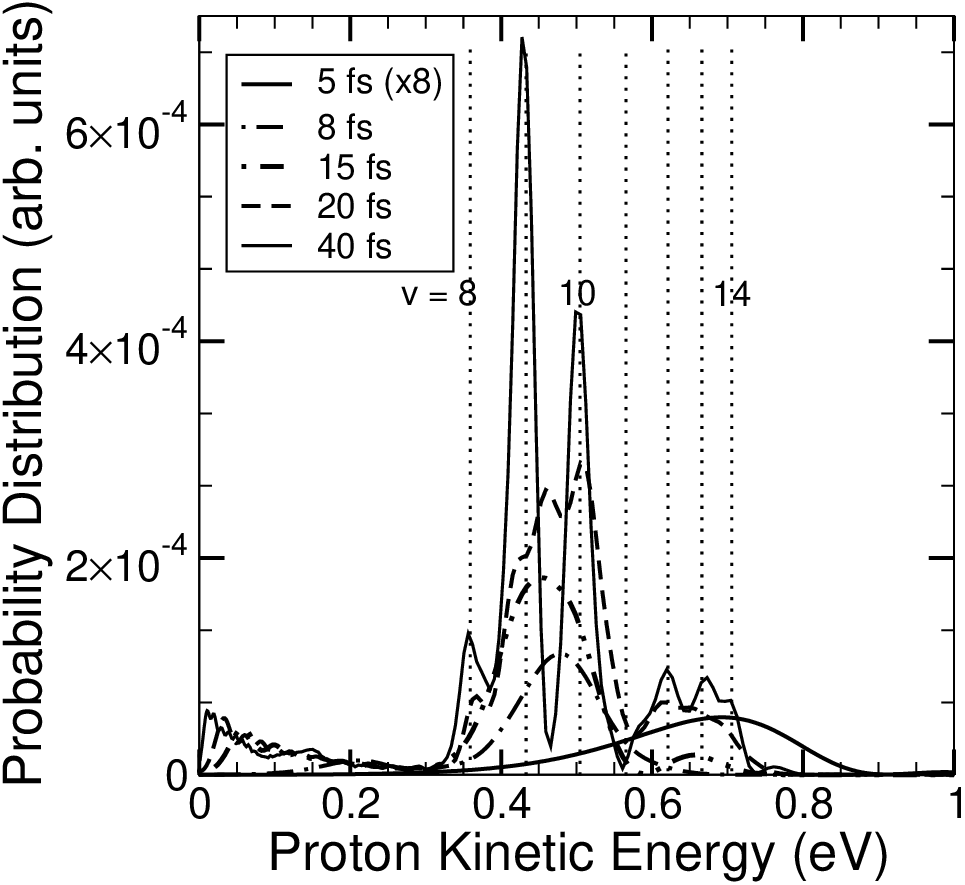}
\caption{Dependence of  proton energy distribution 
on pulse length. I=$5 \times 10^{11}$ W cm$^{-2}$ and
 $\lambda=790$ nm. The corresponding laser pulse length 
is indicated in the figure for each curve.} 
\label{figure5}
\end{figure}

In the well-known 'pump and dump' technique, for heavy 
atoms within a molecule,  the vibrational motion can be easily 
resolved by 100 fs pulses. However for the hydrogen atom the vibration 
is so fast ($\sim 15$fs) that this requires  pulses of 40 fs or 
shorter. We have already shown (figure \ref{figure2}) that  for
 H$_2^+$, the vibrational phase has a role 
 for 20fs or less.   In this section 
we consider the effect on the coherent wavepacket of $v'$ states 
created from the $v=0$ state, and speculate 
on whether this can be resolved by experiments measuring
the proton emission spectrum. 
We begin with a study of pulse length effects for 
the coherent  state. In figure (\ref{figure5}) 
the features of the  low-intensity pulse are considered for $\lambda=790$nm and
$I=5\times 10^{11}$ W cm$^{-2}$,  choosing 
the molecular vibration phases by starting the clock (pulse) 
 at $t'=0$. As noticed previously for $v'=0$, for very 
short pulses the bandwidth broadening and reduced duration leads to 
a flat spectrum with low dissociation yield. Increasing the pulse duration
allows the vibrational structure to be clearly resolved at $T_p=40$fs 
and the familiar Franck-Condon distribution highlighted in 
figure (\ref{figure3}) is apparent. 
Also evident 
in figure(\ref{figure5}) is the interference 
of the shake-off background with the low-energy spectrum.

\begin{figure}
\centering
\includegraphics[clip=true,width=10.0cm]{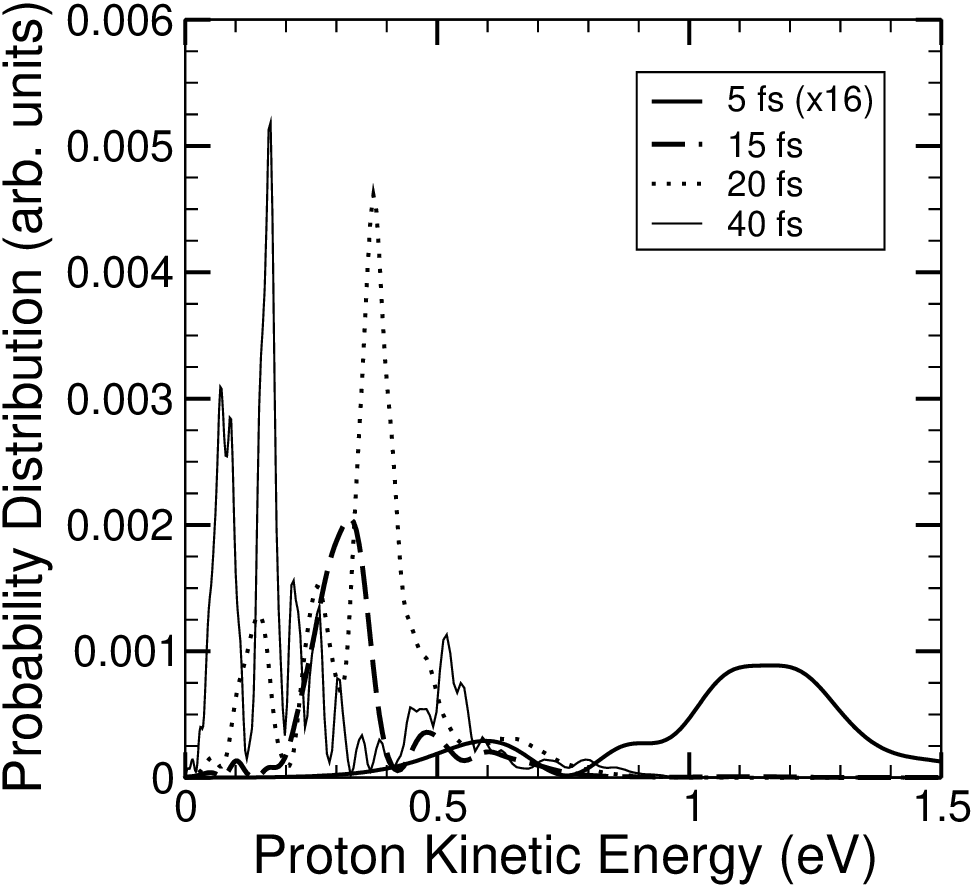}
\caption{Dependence of proton energy distribution on pulse 
length.  $I = 5\times 10^{13}$ W cm$^{-2}$ and 
 $\lambda=790$ nm. The pulse 
duration is indicated in the figure.} 
\label{figure6}
\end{figure}

At higher intensity it has been shown that the vibrational 
 signature in the dissociation spectrum  can be destroyed
to a large extent, even for long pulses. 
Increasing intensity to $I=5\times 10^{13}$ W cm$^{-2}$,
confirms that this occurs for the coherent state (figure \ref{figure6}). Firstly
, the proton energies shift to less than 1eV. We 
note that for the 40 fs pulse, the data is very similar to 
the 55 fs pulse shown in figure \ref{figure4} (c). 
At the other end of the time scale, for  5fs pulses, 
the protons emerge with much higher energies

\begin{figure}
\centering
\includegraphics[clip=true,width=13.5cm]{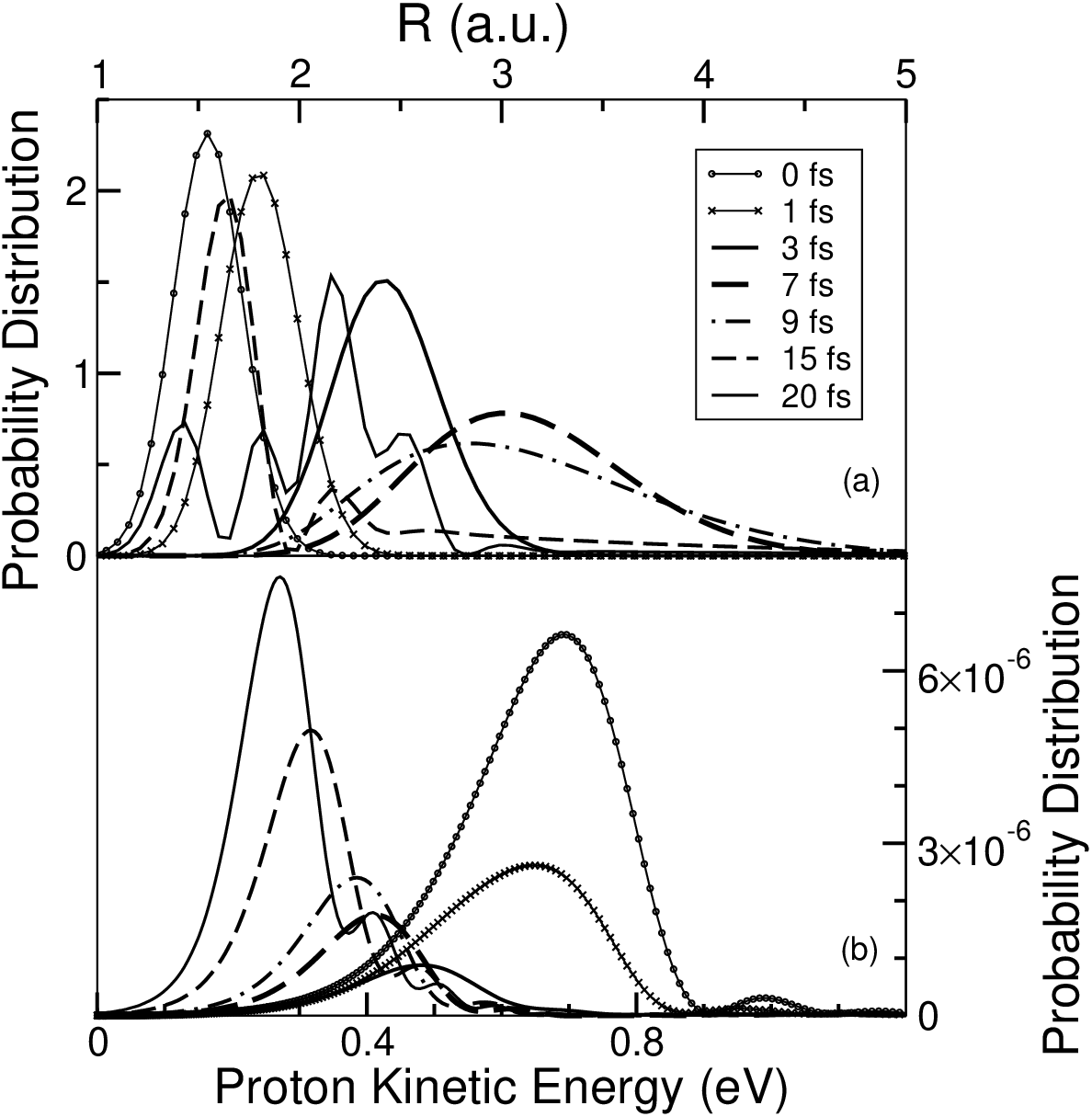}
\caption{Effect of vibration phase (time delay) 
for 5fs pulses. 
$ I=5 \times 10^{11}$ W cm$^{-2}$ and  $\lambda=790$ nm.
(a) The $R$-space probability density at the start of the pulse;
(b) The corresponding proton  energy distribution after the pulse.
The different time delays are indicated in frame (a).} 
\label{figure7}
\end{figure}

That is, apart from the broadening of the peak due to bandwidth 
effects, the wavepacket favours the high frequency 
part of the pulse.  However  it is clear that in this case 
the pulse delay, or molecular phase, has an important role 
\cite{Niik02, Niik03} and \cite{Tong03, Tong04}. In figure (\ref{figure7})
we consider the 5 fs pulse and
we present results for the wavepacket distribution in space 
along with the proton energy spectrum. The pulse is delayed by 
up to 20 fs with respect to the molecular vibration. Firstly
examining the energy spectrum, there is clearly 
a strong variation with pulse delay. 
The 0.7eV peak that appears in 
figure \ref{figure5}  corresponding to the 5 fs peak with 
time delay 0 fs is clearly present in figure \ref{figure7}. 
The corresponding state in $R$-space is the Gaussian 
wavepacket centred at $R=1.4$. The short delay time allows 
the molecule to dissociate from small $R$ values and accelerate 
along the $\Sigma_u$ curve. With a delay of 1 fs the 
wavepacket moves outwards and disperses, this delay severely decreases 
the dissociation yield and attenuates the dissociation 
energy. After 7fs delay the molecular wavepacket is 
at the outer turning point. A pulse applied at this time yields 
a low yield peaked around 0.4 eV.  As the delay time 
increases the vibrational components separate and 
dephase. At 20 fs for example, the wavepacket is irregular and diffuse 
as the high $v'$ components become evident. The dissociation 
from this wavepacket produces a strong low-energy signal at 0.3 eV. 
Comparing with figure \ref{figure4} for example, this energy 
can only be a bandwidth shifted state. That is, the $v=9$ 
state dissociating through a low energy photon. In a real experiment 
the random phases of vibration mean that these details are 
often lost.  
The molecular phase (clock) can be synchronized by  using 
the same laser for photoionization and dissociation \cite{Niik02,Niik03}.
However, the assumption that primary photoionization of H$_2$
is a vertical transition has recently been shown to be 
invalid \cite{Urba04}. This is not too surprising 
since it is generally not a good assumption for
infrared light at high intensity. However it is a reasonable 
model for UV light or electron impact ionization.  

Clearly the sensitivity 
of the proton spectrum is potentially a very useful tool for 
experimental diagnostics. The difficulty still exists 
in that a real laser pulse has temporal and spatial 
variations that can impede the observation of these details.
For  a  long pulse $T_p= 40 $ fs, our calculations show that 
the kinetic energy distribution is not sensitive to the time-delay 
at the same intensity of   $5 \times 10^{11}$ W cm$^{-2}$. At higher intensity 
of $5\times 10^{13}$ W cm$^{-2}$, the spectrum shows only small structure effects 
due to pulse delay. In order to test our model with experiment 
we  conclude our paper with a study in which we take into account 
the experimental parameters and integrate over  the focal volume 
of the pulse and average over molecular orientation.

i
%

\subsection{Comparison with Experiment}
\begin{figure}[t]
\centering
\includegraphics[clip=true,width=10.0cm]{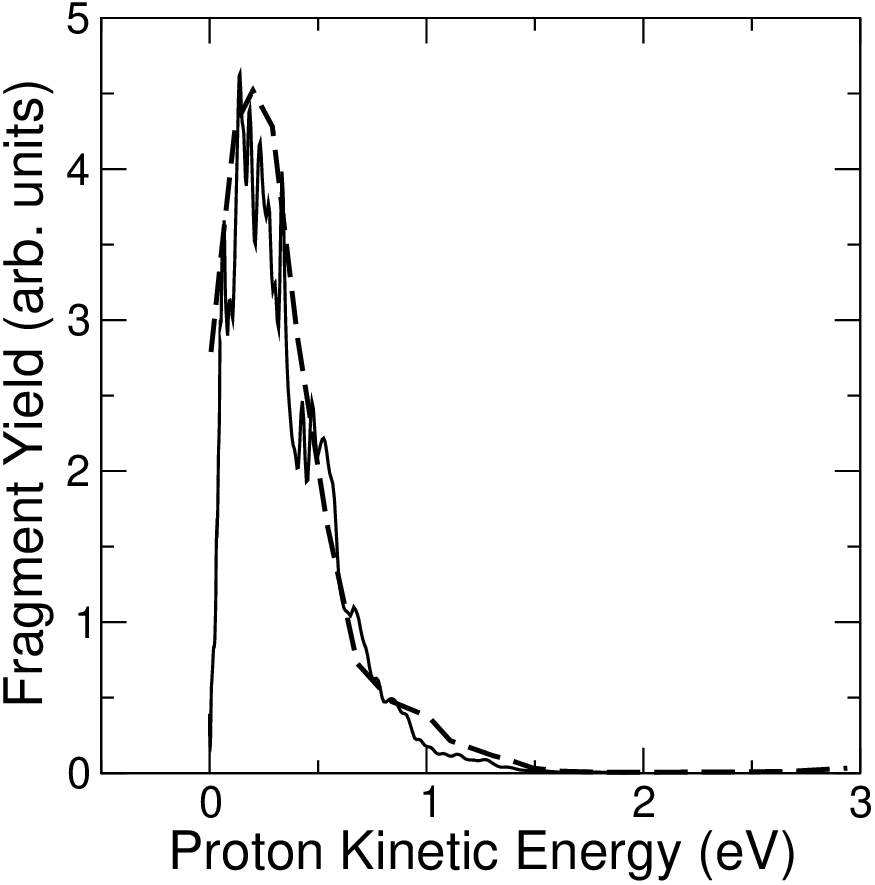}
\caption{Proton  energy distributions from theory and experiment.
$ I=3 \times 10^{15}$ W cm$^{-2}$  
with  $\lambda=790$ nm and $T_p = 65$ fs. $w_0 $ $z_R = $ . 
Full curve: Present calculation; dashed line: experimental
measurement from the pure dissociation channel \cite{Will00}.} 
\label{figure8}
\end{figure}

The experiment \cite{Will00} allows the dissociative and dissociative 
ionization processes can be discriminated by coincidence time-of-flight 
measurements. The spectrum can be dominated by the large low 
intensity focal volume. For a theoretical model the relevant 
spectrum is the probability distribution given by 
\begin{equation}
P(I_f, E_p) =\int \int  2 \pi \rho P_k(I(\rho, z), E_k) d\rho dz
\end{equation}
where $I_f$ is the peak intensity in the focus center and $E_p$ is the kinetic 
energy of the proton. The focus averaging over all these data
with the peak intensity $I_f=2 \times 10^{15}$ W cm$^{-2}$ 
produce figure \ref{figure8} 
which is in excellent agreement with the experiment for the pure 
dissociation peak \cite{Will00}. Unfortunately, 
many of the structures present in the proton energy spectrum 
are lost after the averaging process. The small resonant structures in the theoretical
modelling correspond to the dissociation from different vibrational levels.  
At this intensity, saturation occurs at the centre of the focus, thus a large 
fraction of the dissociation fragments arise from the outer intensity shells.
This is confirmed by our calculation in which the proton spectrum 
has a similar shape for a peak intensity  $5\times 10^{14}$ W cm$^{-2}$ and  
$3 \times 10^{15}$ W cm$^{-2}$. Nonetheless the shape and ion yield of the experiment 
are extremely well characterised in this experiment, and it is  highly encouraging 
that theory and experiment are in such good agreement. More recent measurement 
techniques, develop in the last two years, 
allow sections of the focal volume to be studied, providing and much more 
detailed examination of the proton spectrum. These new measurements 
will hopefully reveal some of the features of the proton 
energy spectrum  described in our paper.


\section{Conclusions}

In summary, we have thoroughly  investigated the dissociation dynamics of 
H$_2^+$ in strong laser fields. The vibrational-state resolved kinetic
energy distributions have been calculated at various intensities ranging from 
$5 \times 10^{9}$ W cm$^{-2}$ to $3 \times 10^{15}$ W cm$^{-2}$ 
for $\lambda =790$nm. The results show the sensitivity of the kinetic
energy distribution on the laser intensity, molecular orientation, 
the pulse length and the pulse 
delay. At very high intensity, our focus-averaging results reproduce
the experimental measurements very accurately \cite{Will00}.  Most 
importantly, we demonstrate that molecular dissociation dynamics at rather 
low intensity with a short pulse duration would be more suitable to act as 
a sub-femtosecond molecular clock. 

\section*{Acknowledgement}
 This research is supported by the award of a PhD research 
studentship from the International Research Centre for Experimental Physics, 
Queen's University Belfast. This work has also been supported  by a grant of 
computer resources at the Computer Services for Academic Research, University 
of Manchester, provided by EPSRC to the UK Multiphoton, Electron Collisions and 
BEC HPC Consortium.

\end{document}